\documentclass[12pt]{article}
\usepackage{amssymb,amsmath}
\def\P1{{\bf P}^1}

\def\0{{\nonumber}}
\def\a{\begin{eqnarray}}
\def\b{\end{eqnarray}}
\def\be{\begin{equation}}
\def\ee{\end{equation}}
\newcommand{\mat}[2][ccccccccccccccccccccccccccccccccccccccccc]{\left(
\begin{array}{#1}
#2\\
\end{array}
\right)}
\begin{document}

\begin{titlepage}

\hfill SISSA/17/2008/EP

\vspace{2.5cm}
 

\centerline{\huge{On gauge/string correspondence and mirror symmetry}}

\vspace{3.0cm}

\centerline{\large{Giulio Bonelli, Houman Safaai}}
\vspace{.5cm}

\centerline{International School of Advanced Studies (SISSA) and INFN, Sezione
  di Trieste}
\centerline{ via Beirut 2-4, 34014 Trieste, Italy}

\vspace{3.5cm}

\begin{abstract}
We consider a mirror dual of the Berkovits-Vafa A-model for the BPS superstring on $AdS_5\times S^5$
in the form of a deformed superconifold. Via geometric transition, the theory has a dual description
as the hermitian gaussian one-matrix model. 
We show that the A-model amplitudes of $AdS_2\times S^4$ branes, breaking
the superconformal symmetry as $U(2,2|4)\to OSp(4^*|4)$, are evaluated in terms of observables
in the matrix model. As such, upon the usual identification $g_{YM}^2=g_s$, these can be expanded as 
Drukker-Gross circular $1/2$-BPS Wilson loops in the perturbative regime of ${\cal N}=4$ SYM.
\end{abstract}

\end{titlepage}

\section{Introduction}

The duality among gauge theories and string theory is a very important subject in modern 
theoretical physics. 
An important issue which is getting much attention is the ability of string theory to reproduce 
known results of the perturbative gauge theory \cite{gopakumar,MHV}. 
In such a regime, the string theory is usually formulated 
in terms of a strongly coupled non linear $\sigma$-model which needs some extra technique to get solved.
This extra technique has been developed in the form of an alternative gauged {\it linear} $\sigma$-model
which reproduces the non linear one in the perturbative regime and 
allows a perturbative expansion in the previously inaccessible regime\cite{phases}.
This program has been fully realized in the case of the three dimensional Chern-Simons theory
by rephrasing its 't Hooft expansion in terms of a topological string on the conifold\cite{ov}.

The case we will try to face here is the string dual of 
a particular sector of the ${\cal N}=4$ SYM in four dimensions, namely
the circular $1/2$ BPS Wilson loops as calculated in \cite{DG} and recently confirmed in \cite{pestun}.

The $AdS_5\times S^5$ string has been shown to admit a formulation in the pure spinor 
framework \cite{B}. In particular it has been shown that to calculate $1/2$-BPS string amplitudes,
one can use a topologically A-twisted version of the ${\cal N}=(2,2)$ $\sigma$-model
on the fermionic coset $U(2,2|4)/U(2,2)\times U(4)$ \cite{ba,BV}. This non-linear $\sigma$-model can be 
obtained by an auxiliary gauged linear one which
has been proposed as the correct framework to describe the string theory in the large curvature regime.

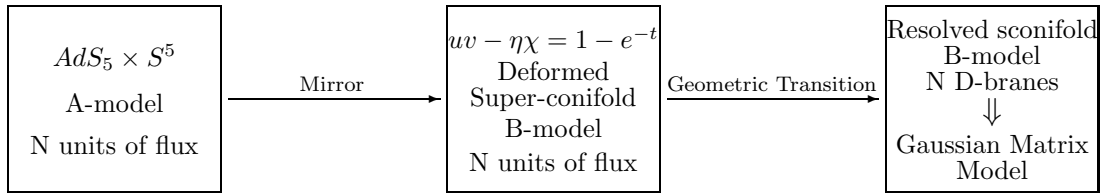
\begin{figure}[hc]
\begin{picture}(40,70)
\put(25,0){\framebox(80,70){\shortstack{\footnotesize$AdS_5\times S^5$\\\\ \\ \footnotesize{A-model}\\ 
\\\\\footnotesize{N units of flux}}}}
\put(108,35){\vector(2,0){80}}
\put(135,38){\scriptsize{Mirror}}
\put(191,0){\framebox(80,70){\shortstack{\footnotesize$uv-\eta\chi=1-e^{-t}$
\\\footnotesize{Deformed}
\\\footnotesize{Super-conifold}
\\\footnotesize{B-model}
\\\\\footnotesize{N units of flux}}}}
\put(274,35){\vector(2,0){80}}
\put(274,38){\scriptsize{Geometric Transition}}
\put(357,0){\framebox(80,70){\shortstack{\footnotesize{Resolved sconifold}\\\footnotesize{B-model}
\\\footnotesize{N D-branes}\\$\Downarrow$\\\footnotesize{Gaussian Matrix}\\\footnotesize{Model}}}}
\end{picture}
\caption{\footnotesize{The duality chain: the mirror symmetry maps to the B-model on the deformed superconifold and 
the geometric transition to the resolved one corresponding to the gaussian matrix model.}}
\label{fig1}
\end{figure}

The aim of this letter is to collect a set of arguments which lead to reproduce the known perturbative 
gauge theory results alluded above by making use of the Berkovits-Vafa proposal \cite{BV}.
Our line of reasoning goes as follows (see Figure \ref{fig1}).

We first consider the A-model for closed strings on $AdS_5\times S^5$ and its gauged linear $\sigma$- model
in the limit of small Fayet-Illiopoulos which corresponds to the large curvature regime. 
In this limit it was noticed already in \cite{BV,CLASS} that the model 
reduces to the invariant quotient $\left(\hat{\bf CP}^{(3|4)}\right)^4// S_4$. Its maximal orbit under the cyclic 
permutation is isomorphic to a single copy of the superprojective space $\hat{\bf CP}^{(3|4)}$.
We can consider then a mirror of such a geometry in the form of a deformed fermionic conifold, dubbed superconifold
\cite{ricci}. This is actually the cotangent bundle over $S^{(1|2)}$ and  
we get the closed B-model with $N$-units of flux along the $S^{(1|2)}$.
We can follow then the theory in a dual formulation after a geometric transition analogous to the Dijkgraaf-Vafa
one \cite{DV, Marino, eiji}. In the superconifold case one calculates the minimal resolution as the resolved superconifold
over $\hat{\bf CP}^{(0|1)}=\left\{{\bf C}^{(1|1)}\setminus (0,0)\right\}/{\bf C}^*$. 
This will be discussed in detail in the main text.
Here the dual theory is that of $N$ D-branes wrapping the base manifold and therefore the theory is described by the
dimensional reduction of the holomorphic $U(N)$ Chern-Simons theory to the branes \cite{Witten1}. 
This results to be the hermitian 
$N\times N$ gaussian  matrix model similar to the purely bosonic case\cite{DV}.
\\
In order to generate gauge invariant observables in the topological string, 
let us now go back to the Berkovits-Vafa $\sigma$-model and look for the A-branes there.
These are wrapped around special lagrangians of the supercoset and their geometry is 
dictated by the possible supersymmetric boundary conditions. On top of the $AdS_4$ branes considered in 
\cite{BV}, there are also other possibilities among which we choose
that of the real 
supercoset $OSp(4^*|4)/SO^*(4)\times USp(4)$. As such, the choice of Dirichlet boundary conditions for 
open strings on such a submanifold breaks the original $U(2,2|4)$ isometry to $OSp(4^*|4)$.
Notice that this is the same symmetry breaking which corresponds to placing 
$1/2$-BPS circular Wilson loops in Minkowski space\footnote{A detailed calculation of this 
can be found in \cite{bianchi}.} as in $\cite{DG}$.
These D-branes can be shown to correspond to $D5$-branes wrapping $AdS_2\times S^4$ geometries
\cite{ST}. 
As such, these states realize the Wilson loops in an alternative way -- suitable for the large curvature regime -- 
compared to the string world-sheet with boundary condition along the loop on the $AdS_5$ boundary.
Analogue constructions were actually elaborated in \cite{electric} (and references therein)
from the point of view of the effective Dirac-Born-Infeld theory, while it is obtained here directly 
for the microscopic theory.

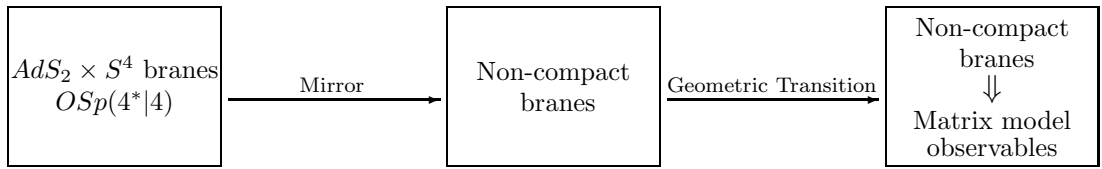
\begin{figure}[hc]
\begin{picture}(40,65)
\put(25,0){\framebox(80,60){\shortstack{\footnotesize{$AdS_2\times S^4$ branes}\\\footnotesize{$OSp(4^*|4)$}}}}
\put(108,25){\vector(2,0){80}}
\put(135,28){\scriptsize{Mirror}}
\put(191,0){\framebox(80,60){\shortstack{\footnotesize{Non-compact}\\\footnotesize{ branes }}}}
\put(274,25){\vector(2,0){80}}
\put(274,28){\scriptsize{Geometric Transition}}
\put(357,0){\framebox(80,60){\shortstack{\footnotesize{Non-compact}\\\footnotesize{ branes }\\$\Downarrow$
\\\footnotesize{Matrix model}\\\footnotesize{observables}}}}
\end{picture}
\caption{\footnotesize{The above duality chain for the $AdS_2\times S^4$-branes. Following them we obtain gaussian matrix 
model amplitudes.}}
\label{fig2}
\end{figure}

We have then to follow these D-branes along the duality map described above (see Figure \ref{fig2}).
Actually the lagrangian cycle is mapped to a transverse non-compact holomorphic cycle in the superconifold
geometry. Therefore, the computation of the corresponding topological string amplitude gets mapped 
to the computation in the gaussian matrix model of the corresponding observables. 
The relevant observables are obtained by integrating over the open strings with mixed 
boundary conditions similar to \cite{BBR}.
\\
This construction therefore leads to express the topological string amplitudes for the A-model on the 
fermionic quotient with $AdS_2\times S^4$-branes boundary conditions as correlators of Wilson loops in the gaussian 
matrix model. As such, these amplitudes should obey the holomorphic anomaly equations of BCOV \cite{BCOV}.
It has been actually proved that it is indeed the case in \cite{EMO}. This not only applies to the construction
in \cite{DV}, but more in general also to the ones given in \cite{adkmv}.
This consistency check strongly supports the validity of our derivation.

The content of this letter is organized as follows.
In the next section we propose a construction of the duality chain leading from the $AdS_5\times S^5$
closed string to the gaussian matrix model.
In the subsequent one we follow the D-branes along the above duality chain and calculate the observables.
The last section is left for consistency checks and few comments.

\section{Strings in $AdS_5\times S^5$ and the mirror geometry}

\subsection{Gauged linear $\sigma$-model of $AdS_5\times S^5$ string theory}

Type IIB String theory on $AdS_5\times S^5$ has been recently formulated 
using pure spinors as a gauged linear $\sigma$-model in \cite{BV}.
It was there shown that the pure spinor IIB superstring action on $AdS_5\times S^5$
can be written up to BRST exact terms as
a nonlinear A-model action defined on a Grassmannian coset whose lowest components 
take values in the supercoset $\frac{U(2,2|4)}{U(2,2)\times U(4)}$
\be
S_{\mathbf{pure}\,\,\mathbf{spinors}}=S_{\mathbf{A-model}}+Q\bar Q X
\ee
Therefore, as far as the calculation of $1/2$-BPS 
observables concerns such an A-model is, upon topological twist, 
equivalent to IIB string theory on $AdS_5\times S^5$. 

The worldsheet variables are fermionic superfields $\Theta_J^A$ and $\bar\Theta_A^J$ where $A=1$ to 4 and $J=1$ to 4 
label fundamental representations of $SU(2,2)$ and $SU(4)$ respectively. 
These $N=2$ chiral superfields can be expanded in components as
\a
\Theta_J^A(\kappa_{+},\kappa_{-})&=&\theta_J^A+\kappa_{+}Z_J^A+\kappa_{-}\bar{Y}_J^A+\kappa_{+}\kappa_{-}f_J^A
\\ \nonumber
\bar\Theta^J_A(\bar\kappa_{+},\bar\kappa_{-})&=&\bar\theta^J_A+\bar\kappa_{+}\bar{Z}^J_A+\bar\kappa_{-}{Y}^J_A+
\bar\kappa_{+}\bar\kappa_{-}\bar{f}^J_A
\b
where $(\kappa_{+},\bar\kappa_{+})$ are left-moving and $(\kappa_{-},\bar\kappa_{-})$ are right-moving Grassmannian 
parameters.

The 32 lowest components $\theta_J^A$ and $\bar\theta_A^J$ are related to the 32 fermionic coordinates of the 
$\frac{PSU(2,2|4)}{SU(2,2)\times U(4)}$ supercoset which parametrizes the $AdS_5\times S^5$ superspace. 
The 32 bosonic variables $Z_J^A$ and $\bar{Z}_A^J$ are twistor-like variables combining the 10 spacetime 
coordinates of $AdS_5$ and $S^5$ with 11 pure spinors $(\lambda_J^A,\bar\lambda^J_A)$ of the pure spinor formalism. 
They can be expressed explicitly as follows
\a
Z_J^A&=&H^A_{A'}(x)(\tilde{H}^{-1}(\tilde{x}))_J^{J'}\lambda_{J'}^{A'}
\\ \nonumber
\bar{Z}^J_A&=&(H^{-1}(x))_{A}^{A'}\tilde{H}_{J'}^J(\tilde{x})\bar\lambda^{J'}_{A'}
\b
where the pure spinors are written in $SO(4,1)\times SO(5)$ notation. 
Here $H^A_{A'}$ is a coset representative for the $AdS_5$ coset $\frac{SU(2,2)}{SO(4,1)}$ 
where $A'=1$ to 4 is an $SO(4,1)$ spinor index and $\tilde{H}_{J'}^J(\tilde{x})$ 
is a coset representative for the $S^5$ coset $\frac{SU(4)}{SO(5)}$ where $J'=1$ 
to 4 is an $SO(5)$ spinor index.
Similarly, the conjugate twistor-like variables $Y_J^A$ and $\bar{Y}^J_A$ 
are constructed from the conjugate momenta to the pure spinors and
$f_J^A$ and $\bar{f}_A^J$ are auxiliary fields.

As discussed in \cite{BV}, the $U(2,2|4)$ invariant action for the topological A-model can be written in the 
mentioned $N=(2,2)$ superfield notation as
\be\label{Amodel}
S=t\int d^2 z \int d^4 \kappa Tr\left[\log(\delta_K^J+\bar\Theta_A^J \Theta_K^A) \right]
\ee
where $t$ is a constant parameter proportional to the $\sigma$-model coupling $R_{AdS_5}^2/\alpha'$.

This A-model is based on a Grassmannian coset $\frac{U(2,2|4)}{U(2,2)\times U(4)}$ and as it was shown already in 
\cite{CLASS}, a nonlinear $\sigma$-model action based on a Grassmannian can be obtained as the Higgs phase of 
an appropriate gauged linear $\sigma$-model.

This is obtained by introducing a $U(4)$ worldsheet gauge field $V_S^R$, together with an 
appropriate set of matter fields transforming in the fundamental representation of the gauge group
\a
\Phi_{R}^\Sigma(z,\bar{z},\kappa^{+},\kappa^{-}),\;\;\;\bar\Phi^{R}_\Sigma(z,\bar{z},\bar\kappa^{+},\bar\kappa^{-})
\b
where $R,S=1$ to 4 are local gauge $U(4)$ indices, and $\Sigma=(A,J)$ is referred to the global $A$ and $J$ indices 
for $U(2,2)$ and $U(4)$ respectively. Note that $\Phi_R^A$ is a fermionic superfield whereas $\Phi_R^J$ is a bosonic 
superfield.
The gauged linear sigma model can be written in $U(2,2|4)$ , $N=(2,2)$ and gauge invariant notation as 
\be
S=\int d^2 z\int d^4 \kappa [\bar\Phi^S_\Sigma \;(e^V)_S^R \Phi_R^\Sigma-t Tr V]
\ee
where $t$ enters as the Fayet-Illiopoulos parameter. When $t$ is nonzero, one can show using the equations of motion 
that the action is equivalent to the $A-$model action (\ref{Amodel}) with the following parametrization for the 
chiral and antichiral superfields $\Theta_J^A$ and $\bar\Theta_A^J$ as follows
\be
\Theta_J^A\equiv \Phi_R^A(\Phi_R^J)^{-1},\;\;\;\bar\Theta^J_A\equiv \bar\Phi^R_A(\bar\Phi^R_J)^{-1}
\ee
As it is noticed in \cite{BV}, in the small $t$ regime, the above gauged linear $\sigma$-model is equivalent
\footnote{Let us notice that here and in the rest of the paper we denote the twistor space
$\hat{\bf CP}^{(m|n)}=\left\{{\bf C}^{(m+1|n)}\setminus\{(0,0)\}\right\}/{\bf C}^*$, where $\{(0,0)\}$ is the origin in
${\bf C}^{(m+1|n)}$.
This space describes the vacua of the gauged linear $\sigma$-model with $m+1$ bosonic chiral multiplets $\Phi_R$
and $n$ fermionic ones $\Phi_A$ all of them with unit charge under the abelian $U(1)$ gauge symmetry.
Its defining equation is $\bar\phi^R\phi_R+\bar\phi^A\phi_A=r$ modulo the $U(1)$ action $\phi_\Sigma\to e^{i\alpha}\phi_\Sigma$. 
We can trade the D-term equation for a complexification of the group action and obtain
the symplectic quotient $\hat{\bf CP}^{(m|n)}$ as defined above. In the mathematical literature, one defines the 
superprojective space ${\bf CP}^{(m|n)}=\left\{{\bf C}^{(m+1|n)}\setminus\left\{{\bf C}^{(0|n)} \right\}\right\}/{\bf C}^*$,
where ${\bf C}^{(0|n)}$ is sitting at the origin $\phi_R=0$ of the commuting variables.
This is a supermanifold contained in $\hat{\bf CP}^{(m|n)}$.
It is clear that the choice of the sublocus containing the origin one has to remove, makes the difference between the two spaces.
The gauged linear $\sigma$-model chooses the sublocus closed under the action of the global $U(m+1|n)$ symmetry of the D-term equations, 
namely the origin of the whole space.
For more formal issues related to supergeometries and all that, see for example \cite{Antonio1} and references therein. 
}, 
by applying an observation at the end of \cite{CLASS}, to the geometric quotient $\left(\hat{\bf CP}^{(3|4)}\right)^4//S_4$. 

For reasons which will be clear in the next section (see the discussion just after (\ref{quotient})), 
let us concentrate on the twisted sector corresponding to the cyclic permutation. 
This is equivalent to a single copy of the twistorial space $\hat{\bf CP}^{(3|4)}$.

\subsection{Mirror symmetry, superconifolds and matrix model}

The first step we need to perform now is a mirror symmetry to relate to the B-model. This has been already calculated in
\cite{AV} and further elaborated in \cite{ricci} for the case at hand.

Let us then consider the A-model on the $\hat{\bf CP}^{(3|4)}$ with bosonic and fermionic coordinates $\phi^I$ and $\phi^A$. 
Since all the fields have charge one under the remnant $U(1)$ gauge group, the D-term equation can be written, in terms 
of the first components of the superfields, as
\be
\sum_{I=1}^4|\phi^I|^2+\sum_{A=1}^4|\phi^A|^2=r
\ee
we can define the dual fields which appear in the mirror theory
\a
Re Y^I&=&|\phi^I|^2
\\ \nonumber
Re X^A&=&-|\phi^A|^2
\b
The superpotential for the mirror Landau-Ginzburg description results to be
\be
\tilde W=\sum_{I=1}^4 e^{-Y^I}+\sum_{A=1}^4 e^{-X^A}(1+\eta^A\chi^A)
\ee
where the fermionic fields $\eta$ and $\chi$ were added to the bosonic field $X$ to match the central charge of
the original $\sigma$-model and to ensure 
the exact matching of the effective superpotentials.
The path integral for the mirror Landau-Ginzburg model can be written as
\be
\label{pathin}
\int \prod_{I=1}^4 dY_I\prod_{A=1}^4 dX_A d\eta_A d\chi_A \delta\left(\sum_{I=1}^4 Y_I-\sum_{A=1}^4 X_A -t \right)
\exp \left( \sum_{I=1}^4 e^{-Y_I}+\sum_{A=1}^4 e^{-X_A}(1+\eta_A\chi_A)\right) 
\ee

This result was further elaborated in \cite{ricci} where it was shown that 
the model has an equivalent mirror picture which is entirely geometric, namely in the form of a superconifold
\be
\int du dv d\eta d\chi dl \exp \left\lbrace l\left(uv-\eta\chi-t'\right)\right\rbrace 
\ee
where $t'=1-e^{-t}$, $\{l,u,v\}$ are bosonic twisted chiral superfields while $\{\eta,\chi\}$ are fermionic 
twisted chiral superfields\footnote{Other mirror pictures were discussed also in \cite{mk}}.

Therefore, as far as the calculation of $1/2$ BPS invariant observables
in Type IIB String theory on $AdS_5\times S^5$ concerns, one can use the mirror 
geometry formulation for the A-model, namely the B-model on the superconifold
\be
uv-\eta\chi=t'
\label{sconifold}\ee
in the regime $t'\sim t\sim 0$.

The geometry in such a regime gets singular. In these situations the string theory target space gets
represented by a blown up geometry via the conifold transition, like in the cases 
which were analyzed in \cite{OV} and \cite{DV}.
One can actually extend the geometric transition to this grassmann odd version of the 
conifold.

The resolved super-conifold is defined by the relations
\be
\mat{u&\eta\\\chi&v}\mat{z \\ \zeta}=0
\ee
where $(z,\zeta)\in \left\{{\bf C}^{(1|1)}\setminus (0,0)\right\}/{\bf C}^*= \hat{\bf CP}^{(0|1)}$.
Away from the singularity it gets mapped to the singular cone $uv-\eta\chi=0$, the singularity
being replaced by $\hat{\bf CP}^{(0|1)}$ very much like in the usual case.
The last space is covered by two patches which we now describe.
If $z\not= 0$, then we can fix our coordinates 
\footnote{Notice that also in the usual bosonic geometric analog, one usually 
specifies the reference points to $z_0=1$, but this is not compulsory at all.}
at any given $z_0\not=0$ as $(z_0,\zeta)$ which is a 
${\bf C}^{(0|1)}$ patch, while if 
$\zeta\not= 0$, then we can fix our coordinates at any given $\zeta_0\not=0$ as $(z,\zeta_0)$ which is a 
${\bf C}^{(1|0)}$ patch. Clearly, on the intersection, the two patches are related by
$z\zeta=z_0\zeta_0$. The last condition is the choice of representative upon the ${\bf C}^*$ equivalent points
exactly as in the usual ${\bf CP}^1$.

Let us now apply the construction of the open string dual theory after geometric transition, 
by following \cite{DV}. This is obtained by realizing the fermionic resolved conifold geometry as a complex structure 
deformation of the local super-$K3$ geometry, namely $\mathcal{O}(-2)\oplus\mathcal{O}(0)$ over $\hat{\bf CP}^{(0|1)}$.
The gluing conditions among the northern and southern hemispheres which are bosonic and fermionic respectively 
are 
\a\label{gluing}
\zeta' z&=&\zeta_0 z_0
\\ \0
\zeta'\psi'&=&z\psi + z_0\phi
\\ \0
\zeta_0 \phi'&=&z_0\phi
\b
where $\psi'$ and $\phi'$ are fermionic while $\psi$ and $\phi$ are bosonic variables. The complex structure deformation 
is induced by the non-diagonal patching term in the second line. Let us call $X$ this superCalabi-Yau space.
The invariant three-form $\Omega$ on $X$ can be defined in this parametrization as follows
\be
\Omega=z_0 d\phi\, d\psi\, dz=\zeta_0\, d\phi' d\psi'\, d\zeta'
\label{Omega}
\ee
in the two coordinate patches.

Similarly to the purely bosonic case, the geometry obtained by imposing the gluing rules can be projected via the blow-down
map

\a
\eta&=&\zeta_0 \psi
\\ \0
\chi&=&z_0 \psi'
\\ \0
u&=&z\psi
\\ \0
x&=&z_0 \phi
\b
which defines the following blown-down geometry
\a
\eta\chi&=&\zeta_0\psi z_0\psi'
\\ \0
&=&\zeta' z \psi \psi'
\\ \0
&=&z\psi(z\psi+z_0\phi)
\\ \0
&=&u(u+x)
\b
which is the singular superconifold (\ref{sconifold}) with $v=u+x$. 

Finally, the resulting matrix model, which is obtained via reduction of the holomorphic Chern-Simons 
theory \cite{Witten1} to the brane, is actually completely analog to the one obtained in the analog bosonic case
\cite{KKLM,DV}.

The open topological B model describing the theory after geometric transition is therefore the reduction 
to the base $\hat{\bf CP}^{(0|1)}$ of 
\be
S=\frac{1}{2g_s}\int_X
\Omega\wedge Tr\left(A\wedge \bar\partial A+\frac{2}{3} A\wedge A\wedge A\right) 
\ee
in the geometry defined in (\ref{gluing}).
For some comments on the Chern-Simons theory on supermanifolds, see also \cite{Antonio2}.
Applying the same reasoning as in \cite{BBR}, one gets
\a
S=\frac{1}{2g_s}\left[ \int_{\hat{\bf CP}^{(0|1)}} Tr(\psi \bar{D}\phi)+
\oint Tr\, W(\phi) \right] 
\b
where $W(x)=\frac{1}{2}x^2$, which reduces to the hermitian gaussian matrix model.
Notice the fact that here, although the base geometry is half fermionic and half bosonic, this does not influence
the endpoint result, because as $\phi$ and $\psi$ change statistics 
while patching, their propagating contributions continue to cancel against the ghost determinants.
The important fact is that the $\bar\partial$-operator on scalars still has a single (constant) zero mode.

Therefore, after geometric transition of the superconifold, one gets 
the gaussian hermitian $N\times N$ matrix model with measure
\be
\mu= dF e^{-\frac{1}{2g_s} Tr F^2}
\ee
which corresponds to the Drukker-Gross one if $g_s=g_{YM}^2$ as predicted by gauge string duality.

\section{D-brane dual observables}

Let us now pass to the discussion of observables in our theory.

We take the boundary conditions for open strings in the coset
$\sigma$-model as follows 
\footnote{Note that these boundary condition are different from the ones which was used in \cite{BV} as 
$(\bar\Theta^t)_J^A=\epsilon^A_{B}{\Theta}^B_K\delta^K_J$. It can be shown that these two type of boundary conditions are producing different types of D-branes.}
\be
(\bar\Theta^t)_J^A=\epsilon^A_{B}{\Theta^*}^B_K\delta^K_J
\label{bc}\ee
where
\footnote{We work in the conventions $\Theta^\dagger=i\bar\Theta$,
$\bar\Theta^\dagger=i\Theta$
and $\left(\psi\zeta\right)^\dagger=-\zeta^\dagger\psi^\dagger$ for fermionic $\psi$ and $\zeta$.}
$\delta$ and $\epsilon$ are
four by four constant matrices such that $\epsilon= a \epsilon^{-1}$
and $\delta= b \delta^{-1}$ with $a$ and $b$ complex numbers such that $ab=-1$.

In order to preserve the correct 1/2 supersymmetry, we chose 
\be
\delta={\mat{1 & 0 \\ 0 & 1}\otimes \mat{0 & -1 \\ 1 & 0}}
\quad {\rm and} \quad
\epsilon={\mat{1 & 0 \\ 0 &1}\otimes \mat{1 & 0 \\ 0 & -1}}
\ee
This breaks the $U(2,2|4)$ isometry to $OSp(4^*|4)$. 

Notice that this remnant symmetry is exactly the same symmetry preserved
by $1/2$ BPS
circular Wilson loops in ${\cal N}=4$ SYM of Drukker and Gross \cite{DG}.

These A-branes wrap the Lagrangian submanifolds of the target space, as
\be
\frac{OSp(4^*|4)}{SO^*(4)\times USp(4)} \longrightarrow
\frac{U(2,2|4)}{U(2,2)\times U(4)}
\ee
which is the fixed locus under the anti-involution 

\be
\bar\Theta\to \delta^t{\Theta^\dagger}\epsilon^t
\quad {\rm and}\quad
\Theta\to {\epsilon^*}^{-1}\bar\Theta^\dagger{\delta^*}^{-1}
\label{reflection}\ee
which is explicitly a symmetry of the $\sigma$-model action since $\delta^{-1}=\delta^\dagger=-\delta$ and $\epsilon^{-1}=\epsilon^\dagger=\epsilon$ in our case.
Recall that $SO^*(4)=SU(1,1)\times SU(2)$ and $USp(4)=SO(5)$ (see \cite{cgmp}).

In the gauged linear $\sigma$- model the boundary conditions (\ref{bc})
become
\be
(\Phi^\dagger)^R_J {\delta^t}^{J}_I= {\kappa^\dagger}^{R}_S \bar\Phi^S_I
\quad{\rm and}\quad
(\Phi^\dagger)^R_A {\epsilon^t}^{A}_B= {\kappa^\dagger}^{R}_S \bar\Phi^S_B
\label{bcg}
\ee
which is the fixed point of the transformation 
\be
\Phi^I_R \to (\delta^\dagger)^I_J (\bar\Phi^\dagger)^J_S \kappa^S_R
\quad{\rm and}\quad
\Phi^A_R \to (\epsilon^\dagger)^A_B(\bar\Phi^\dagger)^B_S\kappa^S_R
\label{transf}
\ee
while $(e^V)\to\kappa e^V\kappa^\dagger$ and
$\kappa$ is,
because of the reality condition on the fields,
a constant element in $O(4)$. 
This breaks the gauge symmetry to ones preserving $\kappa$, namely
$\Lambda\in U(4)$ such that $\Lambda^t\kappa\Lambda=\kappa$.

Actually, upon the reduction to the Coulomb branch
\be
\left(\hat{\bf CP}^{(3|4)}\right)^4 // S_4
\label{quotient}
\ee 
$\kappa$ selects the twisted sector 
to which  the A-branes get coupled. Despite the lack of a manifest target space interpretation, we choose $\kappa$ to be 
the cyclic permutation and we restrict our analysis to this sector of the theory, that is a single copy 
of the twistor space $\hat{\bf CP}^{(3|4)}$.
Under this projection the map (\ref{transf}) becomes the standard anti-involution, under which the K\"ahler form 
is odd, whose fixed locus identifies the lagrangian cycle.

This lagrangian cycle can be traced back in the mirror geometry as in \cite{counting}.
Therefore, applying to the mirror dual at hand, the lagrangian submanifold in $\hat{\bf CP}^{(3|4)}$
gets mapped to the non compact holomorphic cycle
\be
\eta=0\,\, ,\quad
uv-\eta\chi=t'
\label{holomirr}
\ee
in the superconifold mirror picture.
In the singular limit these turn out to be ${\bf C}^{(1|1)}$ non compact branes.
Their fate after geometric transition is to stay non compact, so these are along a fibration on the base 
$\hat{\bf CP}^{(0|1)}$
via a complex curve in the fiber direction which has to compensate the superdimension counting.

Therefore, if in the A-model we add M $D5$-branes, these correspond after the duality to $M$ B-branes 
along the above non-compact cycles.
Now, the open string at hand therefore, on top of the sector of N D-branes along the base, also has the open strings 
connecting them with the dual image of the M $D$-branes.
Correspondingly, the reduced gauge field in the holomorphic Chern-Simons theory becomes
\be
{\cal A}=\mat{A & Y\\ \tilde Y & 0}
\label{extra}
\ee
where the gauge field components $Y$ and $\tilde Y^t$ 
are the $M\times N$ components with mixed boundary conditions.
Being the transverse branes non-compact, the relative gauge field has been kept frozen.
Therefore the action gets reduced as
\be
S_{hCS}({\cal A})=S_{hCS}(A)+\int_X \Omega \wedge Y \bar D_A \tilde Y
\label{intermediate}\ee
where $\bar D_A$ is the covariant $\bar\partial$ operator.

Dimensionally reducing to the base and integrating the reduced $(Y,\tilde Y)$ sector one generates 
the corresponding observable in the matrix model.
In formulas, we have therefore
\be
\int dF e^{-\frac{1}{2g_s}Tr F^2}{\cal O}_M(F)
\ee
By expanding the observable in characters as
\be
{\cal O}_M(F)=\sum_{i,\{n_i\}} {\cal O}_M\left(i,\{n_i\}\right) \prod_i Tr e^{n_i F}
\label{obs}
\ee
one obtains the expansion of the $D5$-brane amplitudes in terms of 1/2 BPS circular Wilson lines
(see Section 4 in \cite{DG}).
The explicit dictionary needs a much deeper elaboration on the specific form of the observables
which will follow from the analysis of the reduced theory on the base of the resolved superconifold.
The prototype of such an analysis for the usual conifold is in \cite{OV}, although to be adapted 
to our case.

\section{Conclusions and Open questions}

In this letter we proposed a dual picture for the calculation of 1/2 BPS open string amplitudes on $AdS_5\times S^5$
with boundary conditions (\ref{bc}) in the large curvature regime. These has been shown to reduce to 
observables in the hermitian gaussian matrix model. Identifying $g_s=g_{YM}^2$, we can interpret
those topological string amplitudes as 1/2 BPS circular Wilson loops.

There are two consistency checks of this result which are independent on the duality chain we formulated.
The first is a symmetry argument, which we already recalled in the paper, that is the fact that 
$AdS_2\times S^4$-branes break exactly the same 1/2 superconformal symmetry as the 1/2 BPS circular Wilson loops do.

The second has to do with the ability of the matrix model to reproduce topological strings amplitudes.
Actually, in order for a candidate set of amplitudes to be compatible with the topological gauge symmetry, 
these have to satisfy the consistency conditions of BCOV \cite{BCOV}, namely the holomorphic anomaly equations.
This is a strict constraint on any dual picture one might find for topological string amplitudes.
The fact that our proposed matrix model passes such a non trivial test is due to the analysis performed in 
\cite{EMO} where this was shown much more in general for the matrix models.
Actually, the $D5$-branes amplitudes then gets reduced to matrix integrals at 
finite $N$. The coinciding genus expansion is consistent for the corresponding 
non local observable insertions which we get in the form
$Tr e^{nF}= \oint \frac{dx}{2\pi i } e^{nx} Tr \frac{1}{F-x}$
which is the natural form of the open string generated observables.
It would be interesting to further elucidate the properties of the specific realization via the gaussian hermitian 
matrix model also in direct comparison with the analysis in \cite{YY}.

It is clear that the reduction of the calculation 
of specific perturbative SYM amplitudes
via a topological string model on the twistor space
$\hat{\bf CP}^{(3|4)}$ 
recalls the duality for MHV amplitudes which started in \cite{MHV}.
The relation with this analysis of what it has been discussed here
could led to a better understanding of the features and limits of 
topological string approach to the string realization of the perturbative 
gauge theory.

The results obtained here are still partial and deserve further investigation.

In particular, we have focused on a particular twisted sector of the string
on the geometric quotient
$\left(\hat{\bf CP}^{(3|4)}\right)^4//S_4$, while the complete theory has all the 
other sectors too.
The SYM dual interpretation of those sectors has to be understood and found.
Also, as we have discussing in section 3, there are different possible choices of BPS boundary conditions 
parametrized by the $\epsilon$ and $\delta$ matrix parameters which are corresponding to different D-brane configurations. These could be used also to produce lower BPS sectors
to be implemented in the gauge/string correspondence as lower BPS Wilson loops \cite{Drukker} which some of them have been described as D-brane configurations. Also one can combine different D-brane configurations to get less supersymmetric objects, an example of which can be obtained by combining the $AdS_4$ boundary conditions in \cite{BV}
and ours which one may generate lower BPS D-branes configurations. 
Moreover,
a precise analysis of the $D5$-branes observables 
(\ref{obs})
has to be performed in order to produce a detailed $D$-branes / circular Wilson loops dictionary.
This analysis passes by the complete reduction to the base of the 
holomorphic Chern-Simons theory on the resolved superconifold.
In particular, this passes by the calculation of the determinant of the relevant $\bar\partial_A$-operator
on supermanifolds. These issues will be discussed in the nearest future.

An interesting issue to study would also be the clarification of how to add non perturbative contributions
in the topological strings to get the instanton corrected version of 1/2 BPS circular Wilson loops \cite{bianchi}
\cite{pestun}.
The gauge amplitude contains, on top of the matrix model integral, also the inverse of the gauge group volume and
an instanton contribution.
The first should be calculated in the complete topological string by the contribution of the pure 
Coulomb phase, very much like as in \cite{ov}. The instanton contribution should be obtained by
including D-instantons in the Berkovits-Vafa context.

As a last comment, let us stress that we conjectured in this letter that the conifold transition 
extends to supergeometries. As such, one should be able to test it for the A-model too, along the lines of \cite{ov,OV}. That is 
one should be able to recast in such a different case, the amplitudes in the Chern-Simons theory on $S^{(1|2)}$
in terms of the gauged linear $\sigma$A-model amplitudes on the resolved superconifold. 
This is another open issue we are letting for future publications.

\vspace{1 cm}
{\bf Acknowledgments}
We thank 
L.F. Alday, 
N. Drukker,
P.A. Grassi,
R. Minasian,
T. Okuda,
S. Pasquetti,
M. Petrini,
A. Tanzini,
for useful discussions. In particular, we thank P.A. Grassi
and A. Tanzini for careful reading the manuscript and their suggestions.
G.B. thanks LPTHE-ParisVI and PMIF-ULB for inspiring hosting during the last stages of this work
and the organizers of the workshop "Spring School/Geometry and Topological field theories"
where our results have been presented.
This research has been supported by MIUR
under the program ``Teoria dei Campi, Superstringhe e Gravit\`a'' and by
the European Commission RTN Program MRTN-CT-2004-005104.


\begin{thebibliography}{99}

\bibitem{gopakumar}
R.~Gopakumar,
  {\it From free fields to AdS,}
  Phys.\ Rev.\  D {\bf 70} (2004) 025009
  [arXiv:hep-th/0308184].

R.~Gopakumar,
  {\it From free fields to AdS. II,}
  Phys.\ Rev.\  D {\bf 70} (2004) 025010
  [arXiv:hep-th/0402063].

R.~Gopakumar,
  {\it From free fields to AdS. III,}
  Phys.\ Rev.\  D {\bf 72} (2005) 066008
  [arXiv:hep-th/0504229].

\bibitem{MHV}
E.~Witten,
  {\it Perturbative gauge theory as a string theory in twistor space,}
  Commun.\ Math.\ Phys.\  {\bf 252}, 189 (2004)
  [arXiv:hep-th/0312171].

A.~Neitzke and C.~Vafa,
{\it N = 2 strings and the twistorial Calabi-Yau,}
  [arXiv:hep-th/0402128].

\bibitem{phases}
  E.~Witten,
 {\it Phases of N = 2 theories in two dimensions,}
  Nucl.\ Phys.\  B {\bf 403} (1993) 159
  [arXiv:hep-th/9301042].

\bibitem{ov}
R.~Gopakumar and C.~Vafa,
 {\it On the gauge theory/geometry correspondence,}
  Adv.\ Theor.\ Math.\ Phys.\  {\bf 3} (1999) 1415
  [arXiv:hep-th/9811131].

H.~Ooguri and C.~Vafa,
{\it Worldsheet derivation of a large N duality,}
  Nucl.\ Phys.\  B {\bf 641} (2002) 3
  [arXiv:hep-th/0205297].

\bibitem{DG}
  N.~Drukker and D.~J.~Gross,
  {\it An exact prediction of N = 4 SUSYM theory for string theory,}
  J.\ Math.\ Phys.\  {\bf 42} (2001) 2896
  [arXiv:hep-th/0010274].

\bibitem{pestun}
  V.~Pestun,
  {\it Localization of gauge theory on a four-sphere and supersymmetric Wilson loops,}
  arXiv:0712.2824 [hep-th].

\bibitem{B}
  N.~Berkovits,
 {\it Super-Poincare covariant quantization of the superstring,}
  JHEP {\bf 0004} (2000) 018
  [arXiv:hep-th/0001035].

  N.~Berkovits,
  {\it Quantum consistency of the superstring in $AdS_5 \times S^5$ background,}
  JHEP {\bf 0503} (2005) 041
  [arXiv:hep-th/0411170].

\bibitem{ba}
N.~Berkovits,
  {\it A New Limit of the $AdS_5$ x $S^5$ Sigma Model,}
  JHEP {\bf 0708} (2007) 011
  [arXiv:hep-th/0703282].

\bibitem{BV}
  N.~Berkovits and C.~Vafa,
 {\it Towards a Worldsheet Derivation of the Maldacena Conjecture,}
  arXiv:0711.1799 [hep-th].

\bibitem{CLASS}
  S.~Cecotti and C.~Vafa,
 {\it On classification of N=2 supersymmetric theories,}
  Commun.\ Math.\ Phys.\  {\bf 158}, 569 (1993)
  [arXiv:hep-th/9211097].

\bibitem{ricci}
  R.~Ricci,
  {\it Super Calabi-Yau's and special Lagrangians,}
  JHEP {\bf 0703}, 048 (2007)
  [arXiv:hep-th/0511284].

\bibitem{KKLM}
S.~Kachru, S.~H.~Katz, A.~E.~Lawrence and J.~McGreevy,
  {\it Open string instantons and superpotentials,}
  Phys.\ Rev.\  D {\bf 62} (2000) 026001
  [arXiv:hep-th/9912151].

\bibitem{DV}
  R.~Dijkgraaf and C.~Vafa,
  {\it Matrix models, topological strings, and supersymmetric gauge theories,}
  Nucl.\ Phys.\ B {\bf 644} (2002) 3
  [arXiv:hep-th/0206255].
  
  R.~Dijkgraaf and C.~Vafa,
  {\it On geometry and matrix models,}
  Nucl.\ Phys.\ B {\bf 644} (2002) 21
  [arXiv:hep-th/0207106].
  
  R.~Dijkgraaf and C.~Vafa,
  {\it A perturbative window into non-perturbative physics,}
  [arXiv:hep-th/0208048].
  
\bibitem{Marino}
  M.~Marino,
  {\it Les Houches lectures on matrix models and topological strings,}
  [arXiv:hep-th/0410165].

\bibitem{eiji}
E.~Konishi,
  {\it Planar Homological Mirror Symmetry,}
  Int.\ J.\ Mod.\ Phys.\  A {\bf 22}, 5351 (2007)
  [arXiv:0707.0387 [hep-th]].

E.~Konishi,
  {\it Planar Homological Mirror Symmetry II,}
  arXiv:0804.1576 [hep-th].

\bibitem{Witten1}
  E.~Witten,
  {\it Chern-Simons gauge theory as a string theory,}
  Prog.\ Math.\  {\bf 133} (1995) 637
  [arXiv:hep-th/9207094].


\bibitem{bianchi}
  M.~Bianchi, M.~B.~Green and S.~Kovacs,
  {\it Instanton corrections to circular Wilson loops in N = 4 supersymmetric Yang-Mills,}
  JHEP {\bf 0204} (2002) 040
  [arXiv:hep-th/0202003].

\bibitem{ST}
  K.~Skenderis and M.~Taylor,
  {\it Branes in AdS and pp-wave spacetimes,}
  JHEP {\bf 0206} (2002) 025
  [arXiv:hep-th/0204054].

\bibitem{electric} 
S.~J.~Rey and J.~T.~Yee, 
{\it Macroscopic strings as heavy quarks in large N gauge theory and anti-de Sitter supergravity,}
 Eur.\ Phys.\ J.\ C {\bf 22} (2001) 379 [arXiv:hep-th/9803001]. 

N.~Drukker and B.~Fiol,
{\it All-genus calculation of Wilson loops using D-branes,}
JHEP {\bf 0502}, 010 (2005) [arXiv:hep-th/0501109]. 

J.~Gomis and F.~Passerini,
  {\it Wilson loops as D3-branes,}
  JHEP {\bf 0701}, 097 (2007)
  [arXiv:hep-th/0612022].

    J.~Gomis and F.~Passerini,
  {\it Holographic Wilson loops,}
  JHEP {\bf 0608}, 074 (2006)
  [arXiv:hep-th/0604007].

  A.~Mikhailov, 
  {\it Special contact Wilson loops,} 
  [arXiv:hep-th/0211229].


\bibitem{BBR}
  G.~Bonelli, L.~Bonora and A.~Ricco,
  {\it Conifold geometries, topological strings and multi-matrix models,}
  Phys.\ Rev.\ D {\bf 72} (2005) 086001
  [arXiv:hep-th/0507224].

\bibitem{BCOV}  
  M.~Bershadsky, S.~Cecotti, H.~Ooguri and C.~Vafa,
 {\it Kodaira-Spencer theory of gravity and exact results for quantum string amplitudes,}
  Commun.\ Math.\ Phys.\  {\bf 165} (1994) 311
  [arXiv:hep-th/9309140].

\bibitem{EMO} 
  B.~Eynard, M.~Marino and N.~Orantin,
  {\it Holomorphic anomaly and matrix models,}
  JHEP {\bf 0706} (2007) 058
  [arXiv:hep-th/0702110].

\bibitem{adkmv}
  M.~Aganagic, R.~Dijkgraaf, A.~Klemm, M.~Marino and C.~Vafa,
  {\it Topological strings and integrable hierarchies,}
  Commun.\ Math.\ Phys.\  {\bf 261} (2006) 451
  [arXiv:hep-th/0312085].

\bibitem{Antonio1}
R. Catenacci, M. Debernardi, P.A. Grassi, D. Matessi
{\it Balanced Superprojective Varieties,}
arXiv:0707.4246 [math-ph]

\bibitem{AV}
  K.~Hori and C.~Vafa,
 {\it Mirror symmetry,}
  [arXiv:hep-th/0002222].
  
  M.~Aganagic and C.~Vafa,
  {\it Mirror symmetry and supermanifolds,}
  [arXiv:hep-th/0403192].

\bibitem{mk}
S.~P.~Kumar and G.~Policastro,
  {\it Strings in twistor superspace and mirror symmetry,}
  Phys.\ Lett.\  B {\bf 619} (2005) 163
  [arXiv:hep-th/0405236].

\bibitem{OV}
H.~Ooguri and C.~Vafa,
  {\it Knot invariants and topological strings,}
  Nucl.\ Phys.\  B {\bf 577} (2000) 419
  [arXiv:hep-th/9912123].



\bibitem{Antonio2}
P.~A.~Grassi and G.~Policastro,
{\it Super-Chern-Simons theory as superstring theory,}
  [arXiv:hep-th/0412272].  

\bibitem{cgmp}
B.~Craps, J.~Gomis, D.~Mateos and A.~Van Proeyen,
 {\it BPS solutions of a D5-brane world volume in a D3-brane background from
  superalgebras,}
  JHEP {\bf 9904} (1999) 004
  [arXiv:hep-th/9901060].

\bibitem{counting}
  K.~Hori, A.~Iqbal and C.~Vafa,
  {\it D-branes and mirror symmetry,}
  [arXiv:hep-th/0005247].
  
  M.~Aganagic and C.~Vafa,
 {\it Mirror symmetry, D-branes and counting holomorphic discs,}
  [arXiv:hep-th/0012041].

\bibitem{YY}
S.~Yamaguchi and S.~T.~Yau,
 {\it Topological string partition functions as polynomials,}
  JHEP {\bf 0407} (2004) 047
  [arXiv:hep-th/0406078].


\bibitem{Drukker}
  N.~Drukker, S.~Giombi, R.~Ricci and D.~Trancanelli,
  {\it Supersymmetric Wilson loops on $S^3$,}
  arXiv:0711.3226 [hep-th].

  N.~Drukker, S.~Giombi, R.~Ricci and D.~Trancanelli,
  {\it On the D3-brane description of some 1/4 BPS Wilson loops,}
  JHEP {\bf 0704} (2007) 008
  [arXiv:hep-th/0612168].

\end{thebibliography}
\end{document}